\documentclass[a4paper,10pt,twoside]{cpc-hepnp}
\usepackage{CJK,upgreek,fancyhdr}
\usepackage{multicol}
\usepackage{multicol}
\usepackage{graphicx}
\usepackage{epstopdf}
\usepackage{booktabs}
\usepackage{amssymb,bm,mathrsfs,bbm,amscd}
\usepackage[tbtags]{amsmath}
\usepackage{lastpage}
\usepackage{xcolor}
\usepackage[T1]{fontenc}
\usepackage{CJK}
\usepackage{xcolor}
\usepackage{amsfonts}
\usepackage{epstopdf}
\usepackage{wrapfig} 
\usepackage{subfigure}
\usepackage{graphicx}  
\usepackage{dcolumn}   
\usepackage{bm}
\usepackage{float}
\newcommand{\be}{\begin{equation}}
\newcommand{\ee}{\end{equation}}
 \newcommand{\bea}{\begin{eqnarray}}
 \newcommand{\ena}{\end{eqnarray}}

\begin{document}
\begin{CJK*}{GB}{gbsn}

\fancyhead[c]{\small Chinese Physics C~~~Vol. xx, No. x (201x) xxxxxx}
\fancyfoot[C]{\small 010201-\thepage}

\footnotetext[0]{Received   }

\title{Thermodynamics and weak cosmic censorship conjecture   of an AdS black hole with a monopole  in the extended phase space\thanks{Supported by National Natural Science
Foundation of China(Grant Nos. 11875095), and Basic Research Project of Science and Technology Committee of Chongqing (Grant No. cstc2018jcyjA2480). }}

\author{Xin-Yun Hu(ºúÜ°ÔÈ)$^{1)}$,
\quad Ke-Jian He(ºÎ¿Æ½¡ )$^{2}$,
\quad Xiao-Xiong Zeng(ÔøÏþÐÛ)$^{3)}$,
\quad Jian-Pin Wu$^{4;1}$(Î⽡Ƹ)\email{keyang@swu.edu.cn}%
}
\maketitle

\address{%
$^1$ College of Economic and Management, Chongqing Jiaotong University, Chongqing 400074, China\\
$^2$ Physics and Space College, China West Normal University, Nanchong 637000, China\\
$^3$ Department of Mechanics, Chongqing Jiaotong University, Chongqing 400074, China \\
$^4$ Center for Gravitation and Cosmology, College of Physical Science and Technology,\\ Yangzhou University, Yangzhou 225009, China\\
}

\begin{abstract}
The first law of black hole thermodynamics  in the extended phase space is prevailing recently. However, the second law as well as  the weak cosmic censorship conjecture has not been investigated extensively. In this paper, we investigate the laws of thermodynamics and the weak cosmic censorship conjecture of an AdS black hole with  a global monopole   in the extended phase space under a charged particle absorption. It is shown that the first law of thermodynamics is  valid, while  the second law   is   violated for the extremal and near-extremal black holes. Moreover, for the weak cosmic censorship conjecture,  we find  it is valid only for the extremal black holes, and it is  violable for the near-extremal black hole, which is different from the previous results.
\end{abstract}



\footnotetext[0]{\hspace*{-3mm}\raisebox{0.3ex}{$\scriptstyle\copyright$}2013
Chinese Physical Society and the Institute of High Energy Physics
of the Chinese Academy of Sciences and the Institute
of Modern Physics of the Chinese Academy of Sciences and IOP Publishing Ltd}%

\begin{multicols}{2}

\section{Introduction}\label{sec:2}
In 1974, Stephen Hawking proved that the black hole had the quantum radiation with the temperature   $T=\frac{\kappa }{2\pi }$ \cite{Hawking:1974sw}. This discovery promoted the investigation on thermodynamics of black holes, and since then, it is believed that a black hole  could be viewed as a thermodynamic system \cite{ref2, Hawking:1976de}. Until now,  there have been many  work on thermodynamics of black holes, such as the four laws of thermodynamics \cite{Bekenstein:1972tm, Bekenstein:1974ax}, quantum effect \cite{Hawking:1974sw,  Zeng:2015nfa, ref018}, phase transition \cite{Zeng:2016sei, Zeng:2016aly}, and so on. In these studies, the AdS space is more popular. In the AdS  space-time,  the cosmological constant can be considered as a thermodynamic variable \cite{Caldarelli:1999xj}. The thermodynamic phase space thus is extended, and  some new thermodynamic phenomena  emerge. So far there have been many researches concentrating on the extend phase space of black holes. Especially, as the cosmological constant and its conjugate quantity were  viewed as the pressure and volume of the black holes, one can construct the extended first law of thermodynamics \cite{ref4}. In addition, in the extended phase space, one also can discuss  the $P-V$ critical behavior \cite{ref4,ref5,ref6}. In this framework, some interesting phenomena, such as Van der Waals   phase transition \cite{ref7,ref8,ref9}, engine cycle \cite{Johnson:2015ekr, Johnson:2017ood, Johnson:2019olt}, can be investigated.

Until now, a lot of researches have been conducted on the first law of thermodynamics  and  phase transition of black holes in the extended phase space. However, the second law of thermodynamics and the weak cosmic censorship conjecture (WCCC) have rarely been reported yet.
On the premise that the first law of thermodynamics is valid, it does not mean that the second law of thermodynamics and the WCCC  still holds. Therefore, it is extremely important and necessary to check the validity of the second  law of thermodynamics and  the WCCC in the extended phase space.

Recently, Ref.\cite{ref19}   investigated the thermodynamics and   WCCC in the extended phase space. Their work was based on the Gedanken experiment in  Ref.\cite{ref10}. After a test particle is absorbed by a black hole, they investigated the change of location of  the event horizon.  As a result, they found that the first  law of thermodynamics and  the WCCC in the extended phase space were always valid for the charged  Reissner-Nordstr\"{o}m AdS black hole, and the second  law of thermodynamics was not valid for the extremal and near-extremal black holes.
   Because there is no general method to prove the WCCC in gravity systems, we  should check it in different space-time backgrounds one by one. Based on the idea  in Ref.\cite{ref19}, thermodynamics and  WCCC in the extended phase space of a series black holes  have  been investigated \cite{ref28, ref29, ref30, ref31, ref32, ref33, ref35, ref36, ref37, ref38, ref39, ref40, Hong:2019yiz}.

   In this paper, we will investigate the thermodynamics and WCCC of an  AdS black hole  with a  global monopole. On one hand, we intend to  discuss the influence of global monopole on the thermodynamics and   WCCC of the black hole. On the other hand, we want to explore  whether  the high order corrections to the mass of the absorbed particle will affect the WCCC. In Ref.\cite{ref19},  the
 WCCC was found to be  valid  in the extended phase space for the extremal and  near-extremal  Reissner Nordstr\"{o}m-AdS  black holes. However, we found that they considered only the first order correction to the mass. As the high order corrections to the mass is considered, whether the conclusion will be changed is worth exploring.  The high order corrections are important for us to investigate the WCCC. Recently Refs. \cite{Wald2018,Sorce:2017dst,Ge:2017vun,An:2017phb} investigated the  WCCC by taking into account the second order variation of the
mass of the black holes,  they found that
 the WCCC was    valid for  the non-extremal black holes, which is different from the previous  cases where  the high order corrections were neglected.   As a result, we find the global monopole has few affect  on the thermodynamic laws. That is, as in \cite{ref19}, the first law is valid, and the second law may be violated. But as the second order to the mass of the absorbed particle is considered, the WCCC is violable, which is different form the result in  \cite{ref19}.

The remainder of this paper is outlined as follows. In section 2, we briefly review the motion of a charged particle in an AdS black holes with a global monopole.  In section 3, we investigate the first and second law of thermodynamics  in the extended phase space. In section 4, we investigate  the WCCC in the extended phase space. Especially, the second order correction to the mass of the particle has been considered.   Section 5 is devoted to our conclusions. In this paper, we will adopt $G = c= \hbar = 1$.

\section{The motion of a charged particle in an  AdS black holes with a global monopole}\label{sec:2}
The spherically symmetric AdS black hole solution with a global monopole can be expressed as \cite{ref24}
\begin{equation}
{ds}^2=-F\left(\tilde{r}\right)d\tilde{t}^2+\frac{1}{F\left(\tilde{r}\right)}d\tilde{r}^2+\tilde{r}^2\left(\text{d$\theta $}^2+\sin ^2\theta  \text{d$\phi $}^2\right),\label{metric1}
\end{equation}
where
\begin{equation}
F\left(\tilde{r}\right)=1-8\pi  \eta _0^2-\frac{2\tilde{m}}{\tilde{r}}+\frac{\tilde{q}^2}{\tilde{r}^2}+\frac{\tilde{r}^2}{l^2},\label{metric}
\end{equation}
in which, $\eta _0$ is a parameter related to symmetry breaking, $\tilde{m}$ and $\tilde{q}$ correspond to mass and charge, $l$ is the radius of AdS space, which relates to the  cosmological parameters  with $\Lambda =-\frac{3}{l^2}$. In this gravitational system, the non-zero electromagnetic four-vector component is
\begin{align}
\tilde{A}_{\tilde{t}}=-\frac{\tilde{q}}{\tilde{r}}. \label{eq2.3}
\end{align}
In order to study  various properties of black holes more easily, we will  introduce the following coordinate transformation
\begin{align}
\tilde{t}=\left(1-8\pi \eta _0^2\right){}^{-1/2}t,   \tilde{r}=\left(1-8\pi \eta _0^2\right){}^{1/2}r,  \label{eq2.4}
\end{align}
and redefine the following physical quantities
\begin{align}
m=\left(1-8\pi \eta _0^2\right){}^{-3/2}\tilde{m},  q=\left(1-8\pi \eta _0^2\right){}^{-1}\tilde{q},  \eta ^2=8\pi \eta _0^2.  \label{eq2.5}
\end{align}
In this way, the metric in Eq.(\ref{metric1}) can be written  as
\begin{align}
{ds}^2=-f(r){dt}^2+\frac{1}{f(r)}{dr}^2+\left(1-\eta ^2\right)r^2\left(\text{d$\theta $}^2+\sin ^2\theta  \text{d$\phi $}^2\right),  \label{eq2.6}
\end{align}
with
\begin{align}
f(r)=1-\frac{2m}{r}+\frac{q^2}{r^2}+\frac{r^2}{l^2}.  \label{eq2.7}
\end{align}
At the same time, the potential of the black hole is
\begin{align}
A_t=-\frac{q}{r}.  \label{eq2.8}
\end{align}
In this case, when $\eta =0$, the black hole returns  to a four-dimensional Reissner-Nordstr\"{o}m AdS  black hole. The existence of the global monopole will also affect the ADM mass and charge of the black hole, namely
\begin{align}
M=\left(1-\eta ^2\right)m,   Q=\left(1-\eta ^2\right)q.  \label{eq2.9}
\end{align}

In this section, we will consider the energy momentum relation of the   charged particle as it is  absorbed by the black hole.  In the potential field $A_\mu$,  the Hamilton-Jacobi  equation can be expressed as
\begin{align}
g^{\mu \nu }\left(p_{\mu }-{eA}_{\mu }\right)\left(p_{\nu }-{eA}_{\nu }\right)+u^2=0, \label{eq2.10}
\end{align}
where $u$ is the mass,  $e$ is the charge, and   $p_{\mu }$  is the momentum of the particle which can be expressed as
\begin{align}
p_{\mu }=\partial _{\mu }\mathcal{I}, \label{eq2.11}
\end{align}
in which $\mathcal{I}$  is the Hamiltonian action of the particle. Considering the symmetry of the gravitational system, this action can be separated into
\begin{align}
\mathcal{I}=-E t+\mathcal{I}_r(r)+\mathcal{I}_{\theta }(\theta )+L \phi, \label{eq2.12}
\end{align}
here, $E$ and $L$  are the energy and angular momentum of the particle. Form   Eq.(\ref{eq2.6}), we can get the inverse metric of the black hole
\begin{align}
g^{\mu \nu }\partial _{\mu }\partial _{\nu }=&-f(r)^{-1}\left(\partial _t\right){}^2+f(r)\left(\partial _r\right){}^2\nonumber\\
&+\left[\left(1-\eta ^2\right)r^2\right]^{-1}\left(\partial _{\theta }{}^2+\sin ^{-2}\theta  \partial _{\phi }{}^2\right). \label{eq2.13}
\end{align}
In this case, the Hamilton-Jacobi  equation can be rewritten as
\begin{align}
u^2-\frac{\left(E+{eA}_t\right){}^2}{f(r)}&+f(r)\left(\partial _r\mathcal{I}_r(r)\right){}^2\nonumber \\
&+\frac{\left(\left(\partial _{\theta }\mathcal{I}_{\theta }\right){}^2+\sin ^{-2}\theta  L^2\right)}{\left(1-\eta ^2\right)r^2}=0. \label{eq2.14}
\end{align}
After  variable separation,   Eq.(\ref{eq2.14}) can be separated into  the following angular and radial equations
\begin{align}
K=\left(\partial _{\theta }\mathcal{I}_{\theta }\right){}^2+\frac{1}{\sin ^2\theta }L^2, \label{eq2.15}
\end{align}
and
\begin{align}
K=&-\left(1-\eta ^2\right)u^2r^2+\frac{\left(1-\eta ^2\right)r^2}{f(r)}\left(-E-{eA}_t\right){}^2\nonumber \\
&-\left(1-\eta ^2\right)r^2f(r)\left(\partial _r\mathcal{I}_r(r)\right){}^2. \label{eq2.16}
\end{align}
In this way, we can rewrite   Eq.(\ref{eq2.12}) as
\begin{align}
\mathcal{I}=\frac{1}{2}m^2\lambda -{Et}+\int {dr}\sqrt{R}+\int {d\theta }\sqrt{\Theta }+{L\phi }, \label{eq2.17}
\end{align}
where
\begin{align}
&\mathcal{I}_r=\int {dr}\sqrt{R}, \quad \nonumber\\
&\mathcal{I}_{\theta }=\int {d\theta }\sqrt{\Theta }, \quad \nonumber\\
&\Theta =K-\frac{1}{\sin ^2\theta }L^2,  \quad \nonumber\\
&R=-\frac{K+\left(1-\eta ^2\right)u^2r^2}{\left(1-\eta ^2\right)r^2f(r)}\nonumber \\
&~~~~~ +\frac{1}{\left(1-\eta ^2\right)r^2f(r)}\left(\frac{\left(1-\eta ^2\right)r^2}{f(r)}\left(-E-{eA}_t\right){}^2\right). \label{eq2.18}
\end{align}
Based on  Eq.(\ref{eq2.17}), the radial and angular momentum can be expressed as
\begin{align}
p^r=f(r)\sqrt{\frac{1}{f^2(r)}\left(E+{eA}_t\right){}^2-\frac{K+\left(1-\eta ^2\right)u^2r^2}{\left(1-\eta ^2\right)r^2f(r)}}, \label{eq2.19}
\end{align}
\begin{align}
p^{\theta }=\frac{1}{\left(1-\eta ^2\right)r^2}\sqrt{K-\frac{1}{\sin ^2\theta }L^2}. \label{eq2.20}
\end{align}
We are most interested in the radial momentum of the particle. At the event horizon, we can get
\begin{align}
E=\frac{q}{r_h}e+\left|p^r_h\right|. \label{eq2.21}
\end{align}
In the front of $\left|p^r_h\right|$, we choose the positive sign. That is,  we want to ensure that particle drops  in the positive direction of time. In this case, the  energy $E$ and momentum $p^r_h$  of the particle are positive too.
\section{Thermodynamic laws in the extended phase space}
\label{sec3}
In order to study the thermodynamics of black holes, we need to obtain some  physical quantities at the event horizon of the black hole. The event horizon of the black hole is defined at $r = r_h$, and its concrete form can be obtained by $f(r_h) = 0$. At the event horizon, the potential energy of the black hole is
\begin{align}
\Phi =\frac{q}{r_h}. \label{eq3.1}
\end{align}
According to the definition of surface gravity, the temperature of the black hole can be expressed as
\begin{align}
T=\frac{\kappa }{2\pi }=\frac{1}{4\pi  r_h}\left[1+\frac{3r_h^2}{l^2}-\frac{Q^2}{\left(1-\eta ^2\right)^2r_h{}^2}\right].  \label{eq3.2}
\end{align}
And the entropy of the black hole can be obtained as
\begin{align}
S=\frac{A_{bh}}{4}=\pi  \left(1-\eta ^2\right)r_h{}^2.  \label{eq3.3}
\end{align}
Recently, some researches have shown that the cosmological parameter could be regarded as the pressure in the thermodynamic system of the black hole, and the corresponded conjugate quantity was viewed as the volume of the system. In this case, the first law of thermodynamics  still holds \cite{ref6}, that is,
\begin{align}
{dM}=T {dS}+\Phi  {dQ}+V {dP}, \label{eq3.4}
\end{align}
where
\begin{align}
P&=-\frac{\Lambda }{8\pi }=\frac{3}{8\pi  l^2},\\
 \quad V&=\frac{4}{3}\pi \left(1-\eta ^2\right)r_h{}^3.  \label{eq3.5}
\end{align}
Accordingly, the Smarr relation in the extended phase space can be obtained as
\begin{align}
M=2(T S-V P)+\Phi  Q.  \label{eq3.6}
\end{align}
In the above equation, $M$ is not the internal energy but enthalpy. The relation  between the enthalpy and internal energy is
\begin{align}
M=U+P V.  \label{eq3.7}
\end{align}
When the black hole absorbs charged particles, it is assumed that energy and charge are conserved. According to the first law of thermodynamics, the energy and charge of the black hole will increase accordingly, which implies
\begin{align}
E=d U=d(M-P V), \quad e=dQ. \label{eq3.8}
\end{align}
Combining the Eq.(\ref{eq3.8}) and   Eq.(\ref{eq2.21}), we can get
\begin{align}
{dU}=\frac{q}{r_h}{dQ}+\left|p^r_h\right|.  \label{eq3.9}
\end{align}
 After the  charged particles are absorbed by black holes, the enthalpy,  charge,   pressure and  volume of the black holes will change accordingly, which are labelled as   $(dM, dQ, dl)$. The other variables can be represented by these variables. Our goal is to obtain the first law of thermodynamics  based on  Eq.(\ref{eq3.9}). Therefore, we will discuss  the change in entropy of the black hole under the absorption of  a charged particle. Form  Eq.(\ref{eq3.3}), we have
\begin{align}
{dS}=2\pi  \left(1-\eta ^2\right)r_h{dr}_h.  \label{eq3.10}
\end{align}
The change of the event horizon is determined by the parameters of the absorbed particle $(e, p_h^r)$. Due to the change of the horizon, the function $f(r)$   determining the location of  the horizon will also change accordingly, namely
\begin{align}
{df}_h=\frac{\partial f_h}{\partial M}{dM}+\frac{\partial f_h}{\partial Q}{dQ}+\frac{\partial f_h}{\partial l}{dl}+\frac{\partial f_h}{\partial r_h}{dr}_h=0, \label{eq3.11}
\end{align}
where we have used the relation $f(r_h)=f(r_h+dr_h)=0$, and
\begin{align}
&\frac{\partial f_h}{\partial M}=\frac{2}{r_h \left(\eta ^2-1\right)}, \quad \nonumber\\
&\frac{\partial f_h}{\partial Q}=\frac{2Q}{\left(\eta ^2-1\right)^2r_h{}^2},\quad \nonumber\\
&\frac{\partial f_h}{\partial l}=-\frac{2r_h{}^2 }{l^3}, \quad \nonumber\\
&\frac{\partial f_h}{\partial r_h}=\frac{2r_h}{l^2}-\frac{2\left(\left(\eta ^2-1\right){Mr}_h+Q^2\right)}{\left(\eta ^2-1\right)^2r_h{}^3}. \label{eq3.12}
\end{align}
In addition, duo to $M$ is enthalpy, the Eq.(\ref{eq3.9}) can be rewritten as
\begin{align}
{dM}-d(P V)=\frac{q}{r_h} {dQ}+p^r_h. \label{eq3.13}
\end{align}
Combining   Eq.(\ref{eq3.11}) and   Eq.(\ref{eq3.13}), we can remove the $dl$ term. Interestingly, we find   $dQ$ and $dM$ are also eliminated. Therefore, we finally get
\begin{align}
{dr}_h=\frac{2l^2p^r_hr_h}{2l^2\left(\left(1-\eta ^2\right)r_h-M\right)+\left(1-\eta ^2\right)r_h{}^3} . \label{eq3.14}
\end{align}
Form  Eqs.(\ref{eq3.3}), (\ref{eq3.5}) and (\ref{eq3.14}), the variation of  entropy and volume of the black hole can be written as the function of particle momentum, which are
\begin{align}
{dS}=\frac{4\left(1-\eta ^2\right)\pi  l^2p^r_hr_h{}^2}{2l^2\left(\left(1-\eta ^2\right)r_h-M\right)+\left(1-\eta ^2\right)r_h{}^3}, \label{eq3.15}
\end{align}
\begin{align}
{dV}=\frac{8\left(1-\eta ^2\right)\pi  l^2p^r_hr_h{}^3}{2l^2\left(\left(1-\eta ^2\right)r_h-M\right)+\left(1-\eta ^2\right)r_h{}^3}. \label{eq3.16}
\end{align}
Combine  Eqs.(\ref{eq3.16}), (\ref{eq3.15}), (\ref{eq3.2}) and (\ref{eq3.3}), we find
\begin{align}
T {dS}-P {dV}=p^r_h. \label{eq3.17}
\end{align}
In this case, the energy momentum relation  in   Eq.(\ref{eq3.9}) becomes as
\begin{align}
d U=\Phi dQ+T dS-PdV. \label{eq3.18}
\end{align}
In the extended phase space, the mass of the black hole has already been defined as enthalpy. We can express the internal energy of  above equation as enthalpy through   Eq.(\ref{eq3.7}), namely
\begin{align}
dM=dU+PdV+VdP. \label{eq3.19}
\end{align}
Substituting  Eq.(\ref{eq3.19}) into   Eq.(\ref{eq3.18}),   we can obtain
\begin{align}
dM=TdS+\Phi dQ+VdP, \label{eq3.20}
\end{align}
which is the same as Eq.(\ref{eq3.4}). Obviously, the first law of thermodynamics    holds after the charged particle is absorbed by the AdS black hole with a global monopole.

Now, we discuss the second law of thermodynamics of  the black hole. It is well known that the black hole entropy never decreases in the clockwise direction. In other words, the  black hole absorbs  the charged particle should be a process of entropy incasement. Thereafter, we will check whether this is true in the extended phase space by  Eq.(\ref{eq3.15}).

We first discuss the case of the extremal black hole. A typical feature of the extremal black hole is that  its temperature vanishes. Based on this fact and Eq.(\ref{eq3.2}), we can get the mass of the extremal black hole. Substituting this mass into  Eq.(\ref{eq3.15}), we obtain
\begin{align}
{dS}_{\text{extreme}}=-\frac{4\pi  l^2p^r_h}{3 r_h}<0. \label{eq3,21}
\end{align}
Obviously, the increasement in the entropy of the black hole is negative, implying the second law of thermodynamics is invalid in this case.

Next, we will discuss the near-extremal black hole. After the values of mass $M$ and  the monopole parameter $\eta$ are  given, we find the value of charge $Q_e$ that satisfies extremal conditions by    numerical approximation method. The condition that the black hole can be established is $Q\leq Q_e$. When  the charge and mass are given, we can find the event horizon $r_h$ of the black hole  by  the equation $f(r_h) = 0$. In this way,  based on  Eq.(\ref{eq3.15}), we can get the variation   of   entropy $dS$. In this paper, we fix  the parameters $M = 0.5$, $l = 1$.  For the case $\eta = 0.1$, we find that the  extremal charge is  $Q_e = 0.462988$. For black holes with different charges, the location of event horizon and the variation of entropy
are listed in  Table 1.
\begin{center}
{\footnotesize{\bf Table 1.} The relation between $dS$, $Q$ and $r_h$ while $\eta=0.1$.\\
\vspace{2mm}
\begin{tabular}{ccc}
\hline
$Q $               &$r_h $             & $dS $         \\
\hline
0.462988     & 0.389008           & $-10.9783$     \\
0.46          & 0.425981           & $-28.2074$     \\
0.455              & 0.449251           & $-121.19 $    \\
0.44              & 0.489249           & $35.1871$     \\
0.42             & 0.522924             & $19.2248$      \\
0.4               & 0.547731           & $15.0991 $     \\
0.3               & 0.62191           & $10.2483 $     \\
0.2               & 0.660347         & $9.15501 $     \\
0.1               & 0.680281           & $8.74138$      \\
\hline
\end{tabular}}
\end{center}
It is obvious that as the black hole charge decreases, the horizon radius gradually increases. And the amount of change in charge and entropy is not a simple monotonic relationship. When the charge approaches to the extremal charge, the value of change in entropy is negative. When the charge is far from the extremal charge, the change of entropy is positive. In the positive region and the negative region, the  entropy  decreases as the  charge decreases.  The near-extreme black hole violates the second law of thermodynamics, and the far-extreme black hole follows the second law of thermodynamics. In Figure 1, we directly give the relationship between the change of the entropy and the event horizon.
\begin{figure}[H]
\centering
\includegraphics[scale=0.7]{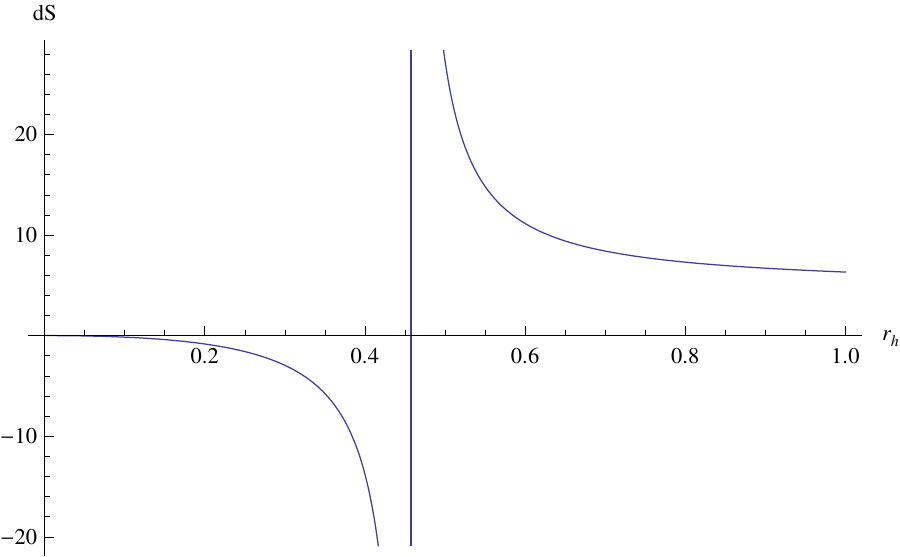}
\caption{The relation between $dS$ and $r_h$ with $M = 0.5, l = p^r_h = 1$ and $\eta=0.1$.}
\label{fig:1}
\end{figure}
Obviously, there is a phase transition point at $r_h = 0.4566$, which is larger than that of the extremal black hole. When the value of the event horizon radius is smaller than this value, the increase in entropy is negative. Conversely, when the value of the horizon radius is greater than this value, the increase of the entropy is positive.  This conclusion is completely consistent with the conclusion given in Table 1.

In addition, for different values of $\eta$, we find that the above conclusions will not change. When $\eta = 0.5$, the extreme charge is $Q_e = 0.448134$. The relationship between the change in entropy and the  charge as well as   the horizon radius are shown in Table 2.
\begin{center}
{\footnotesize{\bf Table 2.} The relation between $dS$, $Q$ and $r_h$ while $\eta=0.1$.\\
\vspace{2mm}
\begin{tabular}{ccc}
\hline
$Q $               &$r_h $             & $dS $         \\
\hline
0.448134     & 0.466004           & $-9.09252$     \\
0.44          & 0.537818           & $-35.5889$     \\
0.43              & 0.571502           & $-1118.76 $    \\
0.42              & 0.595534           & $64.6394$     \\
0.41             & 0.61485             & $36.8817$      \\
0.4               & 0.631209           & $27.7267 $     \\
0.3               & 0.72788           & $13.1042 $     \\
0.2               & 0.775407         & $11.0511 $     \\
0.1               & 0.799693           & $10.3366$      \\
\hline
\end{tabular}}
\end{center}

Form Table 2, we also find that the change in entropy is negative for the case of the near-extremal black hole, and the second law of thermodynamics is violated. For far-extremal black holes, the amount of shift in entropy is positive, and the second law of thermodynamics  holds. The relation between the change in entropy and the event horizon is shown in Figure 2. Similarly, we can find a phase transition  point at $r_h = 0.5727$, and the result shown in Figure 2 is consistent with the result in Table 2.

\begin{figure}[H]
\centering
\includegraphics[scale=0.7]{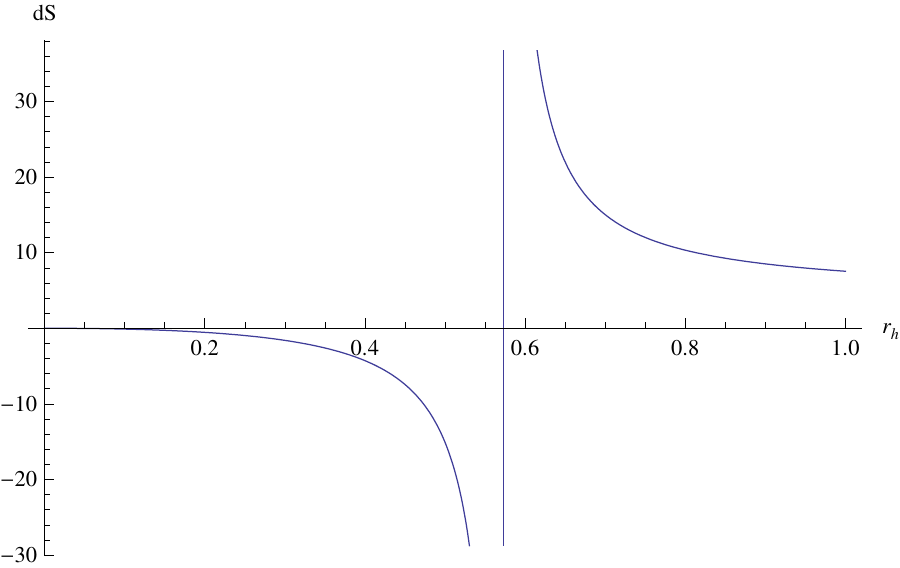}
\caption{The relation between $dS$ and $r_h$  with $M = 0.5, l = p^r_h = 1$ and $\eta=0.5$.}
\label{fig:2}
\end{figure}

\section{The weak cosmic censorship conjecture in the extended phase space}
\label{sec4}
In the extended phase space with consideration of the thermodynamic volume, we find that the second law of thermodynamics is not valid for the case of the extremal black hole and the near-extremal black hole. Therefore, we want to further check whether the WCCC is valid  in the extended phase space.

After a particle is absorbed by the black hole, the change of the black hole is reflected in   $M, Q, l$. However, these physical quantities are ultimately related to the metric function $f(M, Q, l, r)$. Therefore, we mainly discuss the changes in $f(M, Q, l, r)$. For the  function $f(M, Q, l, r)$, there is always a minimum value $r_{\min}$ which satisfies
\begin{align}
f(M,Q,l,r)|_{r=r_\text{min}}\equiv f_\text{min}=\delta\leq 0,
\end{align}
\begin{align}
\partial_{r} f(M,Q,l,r)|_{r=r_\text{min}}\equiv f'_\text{min}=0,  \label{eq4.1}
\end{align}
\begin{align}
 (\partial_{r})^2 f(M,Q,l,r)|_{r=r_\text{min}}>0.  \label{eq4.2}
\end{align}
For the case of the  extremal black hole, $\delta = 0$, and for  the  near-extremal black hole, $\delta$ is a very small negative value. The inner and outer horizons of the black hole are distributed on both sides of $r_{\min}$. After  the  absorption of particles,  $(M, Q, l)$ change into   $(M + dM, Q + dQ, l + dl)$.  Due to these changes, the position of the minimum value of the function $f(M, Q, l, r)$ and the position of the event horizon will move to $r_{\min }\rightarrow r_{\min }+{dr}_{\min }$, $r_h\rightarrow r_h+{dr}_h$. Then, the metric function $f(M, Q, l, r)$ will also have a change, which is labelled  as ${df}_{\min}$. At  $r_{\min }+{dr}_{\min }$, we have
\begin{align}
\partial_{r} f|_{r=r_\text{min}+dr_\text{min}}=f'_\text{min}+df'_\text{min}=0.  \label{eq4.3}
\end{align}
That is
\begin{align}
df'_\text{min}=\frac{\partial f'_\text{min}}{\partial M}dM+\frac{\partial f'_\text{min}}{\partial Q}dQ+\frac{\partial f'_\text{min}}{\partial l}dl+\frac{\partial f'_\text{min}}{\partial r_\text{min}}dr_\text{min}=0,  \label{eq4.4}
\end{align}
 where
\begin{align}
&\frac{\partial f'_{\min }}{\partial M}=\frac{2}{r_{\min }^2\left(1-\eta ^2\right)}, \quad  \frac{\partial f'_{\min }}{\partial Q}=-\frac{4Q}{\left(\eta ^2-1\right)^2r_{\min }^3},  \quad \nonumber\\
&\frac{\partial f'_{\min }}{\partial l}=-\frac{4r_{\min }}{l^3}, \quad \nonumber\\
&\frac{\partial f'_{\min }}{\partial r_{\min }}=\frac{2}{l^2}-\frac{4 M}{\left(1-\eta ^2\right)r_{\min }^3}+\frac{6Q^2}{\left(\eta ^2-1\right)^2r_{\min }^4}.  \label{eq4.5}
\end{align}
At $ r_{\min} + {dr}_{\min}$, the  function $f(M, Q, l,r)$ becomes as
\begin{align}
&f(M+dM,Q+dQ,l+dl,r)|_{r=r_\text{min}+dr_\text{min}}=f_\text{min}+df_\text{min}\nonumber\\
&=\delta+\left(\frac{\partial f_\text{min}}{\partial M}dM+\frac{\partial f_\text{min}}{\partial Q}dQ+\frac{\partial f_\text{min}}{\partial l}dl\right).  \label{eq4.6}
\end{align}
In the above equation, the condition $f'_{\min} = 0$ has been used. The most important step is how to give the specific form of the Eq.(\ref{eq4.6}). For the extremal black hole, we have $r_h=r_{\min}$, which means that the Eq.(\ref{eq3.13}) can be used. According to the condition $f'_{\min} = 0$, we can get an  equation about $M$. So that we can obtain lastly
\begin{align}
{dM}&=\frac{l^2Q^2+3\left(\eta ^2-1\right)^2r_{\min }^4}{\left(\eta ^2-1\right)l^2r_{\min }^2}{dr}_{\min }\nonumber \\
&-\frac{2\left({dl}\left(\eta ^2-1\right)^2r_{\min }^5+l^3Q r_{\min }{dQ}\right)}{\left(\eta ^2-1\right)l^3r_{\min }^2}. \label{eq4.7}
\end{align}
Form Eqs.(\ref{eq3.13}) and (\ref{eq4.7}), we find
\begin{align}
df_{\min }=0.  \label{eq4.8}
\end{align}
Therefore, we finally get
\begin{align}
f(M+dM,Q+dQ,l+dl,r)&|_{r=r_\text{min}+dr_\text{min}}=0. \label{eq4.9}
\end{align}
The result shows that the   minimum value of function $f(r)$ has not been changed in the extended phase space. That is, as a  charged particle is absorbed by the extremal black hole, the configuration of the black hole does not change. In other words, the extremal black hole is still an extremal black hole.

We also can discuss the case of the near-extremal black hole. For the near-extremal black hole,  Eq.(\ref{eq3.13}) can not be used since it is valid only at the horizon. But we can expand it  at $r_\text{min }$ with  the relation $r_h=r_\text{min }+\epsilon $, that is
\begin{align}
{dM}&=\frac{\left( r_{\min }{}^5 \left(\eta ^2-1\right)^2 {dl}-l^3 Q r_{\min } {dQ} \right)}{ l^3 r_{\min }^2 \left(\eta ^2-1\right)} \quad \nonumber\\
&+\frac{l^2 \left(Q^2-r_{\min }{}^2 \left(\eta ^2-1\right)^2\right)-3 r_{\min }{}^4 \left(\eta ^2-1\right)^2}{ l^2r_{\min }^2 \left(\eta ^2-1\right)} \quad \nonumber\\
&+\frac{ l^3 Q r_{\min } {dQ}+3 r_{\min }{}^5{dl}+6 l r_{\min }{}^4 \eta ^2{dr}_{\min }+3{  }r_{\min }{}^5 \eta ^4{dl}  }{l^3 r_{\min }{}^3 \left(\eta ^2-1\right)\epsilon^{-1}} \quad \nonumber\\
&-\frac{ l^3 Q^2{dr}_{\min }+3{  }l r_{\min }{}^4{dr}_{\min }+6{  }r_{\min }{}^5 \eta ^2{dl}+3{  }l r_{\min }{}^4 \eta ^4{dr}_{\min }  }{l^3 r_{\min }{}^3 \left(\eta ^2-1\right) \epsilon^{-1}}\nonumber\\
&+\mathcal O(\epsilon )^2.  \label{eq4.10}
\end{align}
Substituting (\ref{eq4.10}) into (\ref{eq4.6}), we have
\begin{align}
&{df}_{\min }=\frac{ \left(l^2 \left(Q^2-r_{\min }{}^2 \left(\eta ^2-1\right)^2\right)-3 r_{\min }{}^4 \left(\eta ^2-1\right)^2\right){dr}_{\min }}{l^2 r_{\min }{}^3 \left(\eta ^2-1\right)^2}\quad \nonumber\\
&~~~~~+\frac{2 \left({dQ} l^3 Q r_{\min }+3{dl} r_{\min }{}^5 \left(\eta ^2-1\right)^2\right) \epsilon }{l^3 r_{\min }{}^4 \left(\eta ^2-1\right)^2} \quad \nonumber\\
&~~~~~-\frac{2 {dr}_{\min } l \left(l^2 Q^2+3 r_{\min }{}^4 \left(\eta ^2-1\right)^2\right)\epsilon }{l^3 r_{\min }{}^4 \left(\eta ^2-1\right)^2} \quad \nonumber\\
&~~~~~+\mathcal O(\epsilon )^2.  \label{eq4.11}
\end{align}
At $r_h=r_{\min}+\epsilon$, we also can get
\begin{align}
Q=\frac{r_{\min } \sqrt{l^2+3 r_{\min }{}^2} \left(\eta ^2-1\right)}{l},  \label{eq4.12}
\end{align}
and
\begin{align}
{dQ}=\frac{\left(\left(l^3+6 l r_{\min }{}^2\right){dr}_{\min }-3 r_{\min }{}^3{dl}\right) \left(\eta ^2-1\right)}{l^2 \sqrt{l^2+3 r_{\min }{}^2}}.  \label{eq4.13}
\end{align}
Substituting Eqs.(\ref{eq4.12}) and (\ref{eq4.13}) into the Eq.(\ref{eq4.11}), we find
\begin{align}
df_{\min }=\mathcal O(\epsilon^2 ),  \label{eq4.14}
\end{align}
then, we can get
\begin{align}
f_\text{min}+df_\text{min}=\delta+\mathcal O(\epsilon^2 ). \label{eq4.15}
\end{align}
 In  Ref.\cite{ref19}, it was claimed that $df_{\min }$ can be  neglected for  $ \mathcal O(\epsilon^2 )$ is a  small quantity. However,  $\delta$  is also a small  negative quantity, we can not directly determine which is smaller. Therefore, for the near-extremal black hole in the extended phase space, we  should find the relationship between  $\delta$ and $\mathcal O(\epsilon^2 )$. Expanding $f(r_h)$ to the second order, we have
\begin{align}
f\left(r_{\min }+\epsilon \right)=\delta +\frac{\left(l^2 Q^2+3 r_{\min }{}^4 \left(\eta ^2-1\right)^2\right)\epsilon ^2}{l^2 r_{\min }{}^4 \left(\eta ^2-1\right)^2}+\mathcal O(\epsilon^3 )=0, \label{eq4.18}
\end{align}
in which, we have used the relation $r_h=r_\text{min }+\epsilon $, so we can obtain
\begin{align}
\delta=-\frac{\left(l^2 Q^2+3 r_{\min }{}^4 \left(\eta ^2-1\right)^2\right)\epsilon ^2}{l^2 r_{\min }{}^4 \left(\eta ^2-1\right)^2}-\mathcal O(\epsilon^3 ). \label{eq4.19}
\end{align}
Similarly, expanding Eq.(\ref{eq3.13}) to the second order, and inserting the expanded result into Eq.(\ref{eq4.6}), we can obtain
\begin{align}
{df}_{\min }=\left(\frac{6 (1+{dl})}{l^3}+\frac{2 d r_{\min }}{ r_{\min }{}^3}\right) \epsilon ^2+\mathcal O(\epsilon^3 ). \label{eq4.20}
\end{align}
For simplicity, we  define.
\begin{align}
\Delta _E=\frac{\delta  +{df}_{\min }}{\epsilon ^2}. \label{eq4.21}
\end{align}
Combining the Eqs.(\ref{eq4.19}), (\ref{eq4.20}) and (\ref{eq4.21}), we have
\begin{align}
\Delta _E=\frac{6 (1+{dl})}{l^3}-\frac{3}{l^2}+\frac{2\text{  }r_{\min } d r_{\min }-Q^2\left(\eta ^2-1\right)^{-2}}{ r_{\min }{}^4}.\label{eq4.22}
\end{align}
Obviously,  the value of $\Delta _E$ is directly related to the   values of $(Q, \eta, r_{\min}, l, dl)$.  In order to make it is easier to judge the positive or negative value of $\Delta _E$, we have plotted  Figure 3 and Figure 4 for different values of $(Q, \eta, r_{\min}, l, dl)$. For the case  $\Delta _E$ is positive, there does not exist horizon since the equation $f(r)=0$ does not have solutions. And for the   $\Delta _E$ is negative, there   exist horizons always.  In other words, the  WCCC is violated for the case $\Delta _E>0$.
\begin{figure}[H]
\centering
\includegraphics[scale=0.7]{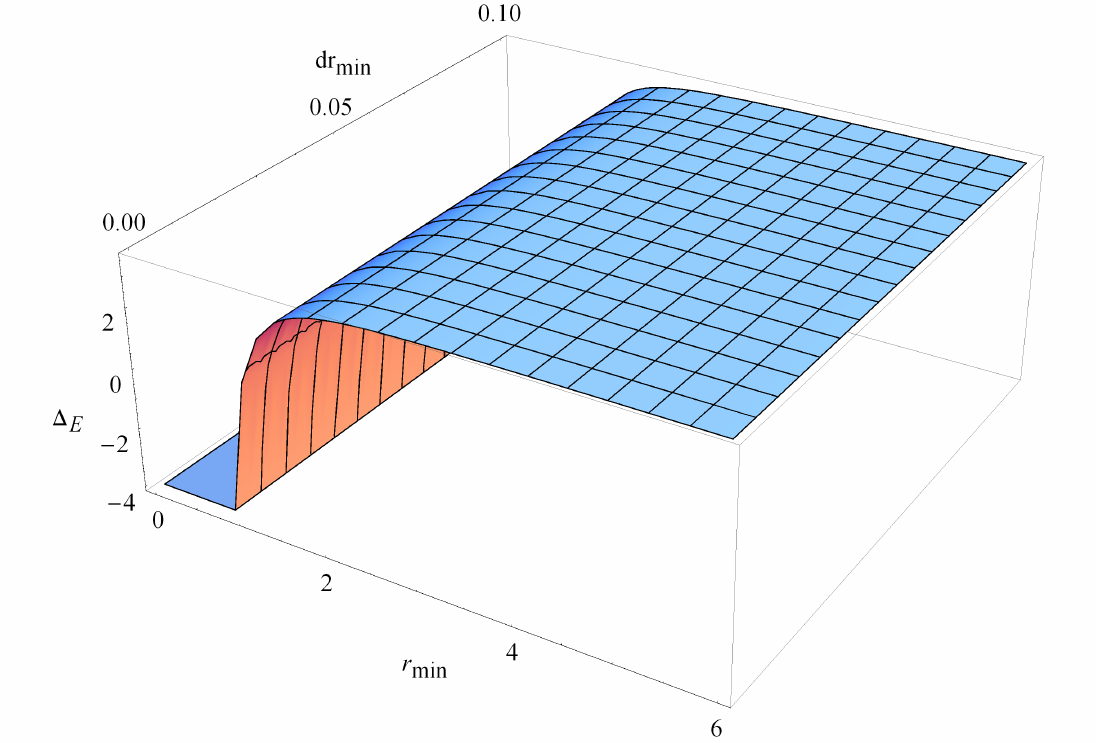}
\caption{The value of  $\Delta _E$ for $l=1, Q=2, \eta=0.1, dl=0.1$.}
\label{fig:3}
\end{figure}

\begin{figure}[H]
\centering
\includegraphics[scale=0.7]{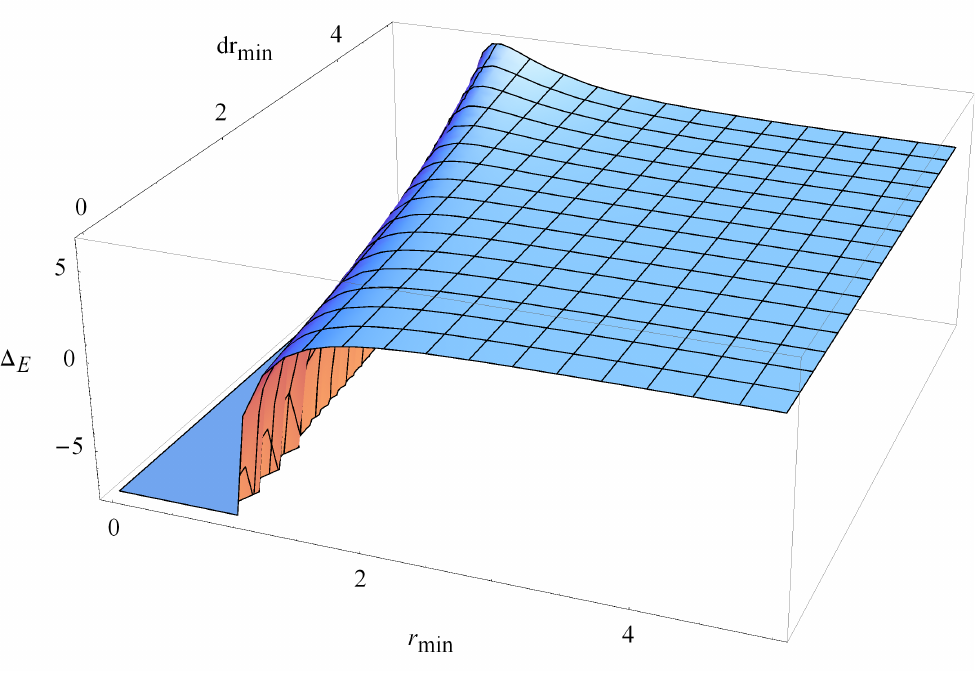}
\caption{The value of  $\Delta _E$ for $l=1, Q=2, \eta=0.5, dl=0.1$.}
\label{fig:4}
\end{figure}
From Figure 3 and Figure 4, we find for different values of  $(Q, \eta, r_{\min}, l, dl)$,  $\Delta _E$ may be positive or negative.   That is, in the extended phase space,  the WCCC is violable for the near-extremal black hole. This  result is different with that in Ref.\cite{ref19}, where the WCCC is valid always. Our results  is more precise and comprehensive since we consider the high order corrections to the mass of the absorbed particles.

\section{Discussion and conclusions}
\label{sec:5}

The thermodynamics of black holes provides an effective means for studying the relationship between the gravity, thermodynamics and  quantum theory. And in-depth study of the thermodynamics of black holes is helpful for us  to further understand the nature of the gravity. In the extended phase space, the laws of thermodynamics   and WCCC were researched under charged particles absorbtion. Based on the Hamiltonian-Jacobian equation, we first got the relation between particle momentum and energy. Then, from this  relation, we  obtained the first law of thermodynamics in the extended phase space, which was found to be   valid. Through the  methods of the numerical analysis, we got the value of the variation of the entropy  in the case of the extremal black hole, near-extremal black holes, and far-extremal black hole. The results shown that for the extremal black hole and the near-extremal black hole, the change of the black hole entropy was negative, while for the far-extremal black hole, it was positive. In other words, the second law of thermodynamics of the black hole was only valid for the far-extremal black hole.

In the extended phase space, we also checked the WCCC. We mainly studied the change  of the minimum value of the function $f(r )$  which determines location of the event horizon. For the case of the extremal black hole, the result shown that $f(r_{\min})$ did not change as a charged particle is absorbed. Therefore, the extremal black hole was always  extremal.   The WCCC hence  was still valid for the extremal black hole. However, we found  that $f(r_{\min}+dr_{\min})>0$ could occur for the case of near-extremal black hole. Thus, the WCCC is  invalid for the near-extremal black hole. This result was quite different  from the result in  Ref.\cite{ref19}. Because we did not neglect the contribution of $\delta$ and $\mathcal O(\epsilon^2 )$ by considering the second order correction to the mass of the particle. It seems that our conclusion  is  more precise and comprehensive.

\end{multicols}

\vspace{10mm}

\vspace{-1mm}
\centerline{\rule{80mm}{0.1pt}}
\vspace{2mm}

\begin{multicols}{2}

\end{multicols}

\clearpage
\end{CJK*}
\end{document}